\documentclass[sigconf]{acmart}

\copyrightyear{2020}
\acmYear{2020}
\setcopyright{acmlicensed}
\acmConference[CIKM '20]{Proceedings of the 29th ACM International Conference on Information and Knowledge Management}{October 19--23, 2020}{Virtual Event, Ireland}
\acmBooktitle{Proceedings of the 29th ACM International Conference on Information and Knowledge Management (CIKM '20), October 19--23, 2020, Virtual Event, Ireland}
\acmPrice{15.00}
\acmDOI{10.1145/3340531.3411962}
\acmISBN{978-1-4503-6859-9/20/10}

\usepackage{fdiaz}

\usepackage{booktabs} 
\usepackage{xspace} 
\usepackage{subcaption}
\usepackage{multirow}
\usepackage{balance}

\newcommand{\queries}[0]{\fdset{Q}}
\newcommand{\query}[0]{q}

\newcommand{\docs}[0]{\fdset{D}}
\newcommand{\doc}[0]{d}
\newcommand{\rdoc}[0]{d^*}
\newcommand{\ndocs}[0]{n}
\newcommand{\nreldocs}[0]{m}
\newcommand{\nnreldocs}[0]{(\ndocs-\nreldocs)}

\newcommand{\permutationset}[0]{S_{\ndocs}}
\newcommand{\permutation}[0]{\sigma}
\newcommand{\invpermutation}[0]{\overline{\sigma}}

\newcommand{\ranker}[0]{\pi}
\newcommand{\scorerparams}[0]{\theta}
\newcommand{\scorer}[0]{f_\scorerparams}

\newcommand{\temp}[0]{\tau}

\newcommand{\qrels}[0]{\fdvec{y}^*}
\newcommand{\iqrels}[0]{\fdvec{y}^{\intent,*}}
\newcommand{\score}[0]{\fdvec{y}}

\newcommand{\nrel}[0]{m}
\newcommand{\grade}[0]{g}

\newcommand{\numsamplestrain}[0]{n_\text{train}}
\newcommand{\numsamplestest}[0]{n_\text{test}}

\newcommand{\metric}[0]{\mu}

\newcommand{\eeLoss}[0]{\ell}
\newcommand{\eeRelevance}[0]{\text{EE-R}}
\newcommand{\eeDisparity}[0]{\text{EE-D}}
\newcommand{\eeAUC}[0]{\text{EE-AUC}}

\newcommand{\rbp}[0]{\text{RBP}}
\newcommand{\rbpp}[0]{\gamma}
\newcommand{\rbpd}[0]{k}

\newcommand{\err}[0]{\text{ERR}}
\newcommand{\errp}[0]{\gamma}
\newcommand{\errd}[0]{k}
\newcommand{\errphi}[0]{\phi}

\newcommand{\ia}[0]{\text{IA-$\metric$}}
\newcommand{\iaRBP}[0]{\text{IA-RBP}}

\newcommand{\exposure}[0]{\epsilon}
\newcommand{\targetexposure}[0]{\exposure^*}
\newcommand{\iexposure}[0]{\xi}
\newcommand{\targetiexposure}[0]{\iexposure^*}

\newcommand{\exposureMassConstant}[0]{c}

\newcommand{\intents}[0]{\fdset{A}}
\newcommand{\nintents}[0]{k}
\newcommand{\intent}[0]{a}
\newcommand{\intentMatrix}[0]{\fdmat{\tilde{A}}}
\newcommand{\attributeMatrix}[0]{\fdmat{A}}

\newcommand{\rwRestartProbability}[0]{\theta}
\newcommand{\rwNumSteps}[0]{k}

\newcommand{\parityreltradeoff}[0]{\lambda}

\newcommand{\instancelossmod}[0]{\ell_\parityreltradeoff}
\newcommand{\instancelossmodgroup}[0]{\ell_{\text{group},\parityreltradeoff}}

\newcommand{\prob}[0]{p}
\newcommand{\probnoise}[0]{\tilde{\prob}}
\newcommand{\gumbelnoise}[0]{G}

\newcommand{\fdexp}[1]{\exp\left({#1}\right)}

\newcommand{\robust}[0]{{\sc Robust2004}\xspace}

\newcommand{\ml}[0]{{\sc MovieLens25M}\xspace}

\begin{document}
\fancyhead{}

\title[Evaluating Stochastic Rankings with Expected Exposure]{Evaluating Stochastic Rankings with Expected Exposure}

\author{Fernando Diaz}
\affiliation{
  \institution{Microsoft}
  \city{Montr\'eal}
  \state{QC}
}
\email{diazf@acm.org}
\author{Bhaskar Mitra}
\affiliation{
  \institution{Microsoft}
  \city{Montr\'eal}
  \state{QC}
}
\email{bmitra@microsoft.com}
\author{Michael D. Ekstrand}
\affiliation{
  \department{People \& Information Research Team}
  \institution{Boise State Computer Science}
  \city{Boise}
  \state{ID}
}
\email{michaelekstrand@boisestate.edu}
\author{Asia J. Biega}
\affiliation{
  \institution{Microsoft}
  \city{Montr\'eal}
  \state{QC}
}
\email{asia.biega@microsoft.com}
\author{Ben Carterette}
\affiliation{
  \institution{Spotify}
  \city{New York}
  \state{NY}
}
\email{carteret@acm.org}

\begin{abstract}
We introduce the concept of \emph{expected exposure} as the average attention ranked items receive from users over repeated samples of the same query.  
Furthermore, we advocate for the adoption of the principle of equal expected exposure: given a fixed information need, no item should receive more or less expected exposure than any other item of the same relevance grade.  
We argue that this principle is desirable for many retrieval objectives and scenarios, including topical diversity and fair ranking.  
Leveraging user models from existing retrieval metrics, we propose a general evaluation methodology based on expected exposure and draw connections to related metrics in information retrieval evaluation.  
Importantly, this methodology relaxes classic information retrieval assumptions, allowing a system, in response to a query, to produce a \emph{distribution over rankings} instead of a single fixed ranking. 
We study the behavior of the expected exposure metric and stochastic rankers across a variety of information access conditions, including \emph{ad hoc} retrieval and recommendation.  
We believe that measuring and optimizing expected exposure metrics using randomization opens a new area for retrieval algorithm development and progress.  
\end{abstract}



\maketitle
\section{Introduction}
\label{sec:introduction}

Information access systems such as retrieval and recommendation systems often respond to an information need with a ranking of items.  
Even with more sophisticated information display modalities, the ranked list is a central feature of most interfaces.
Since users often inspect a ranked list in a nonrandom--usually linear--order, some items are exposed to the user before others.
Even if a system can perfectly model relevance and rank items accordingly, it still must put items in a particular order, breaking relevance ties in some way and reifying small differences in relative relevance into distinct rank positions.

\begin{figure}[t]
\includegraphics[width=2in]{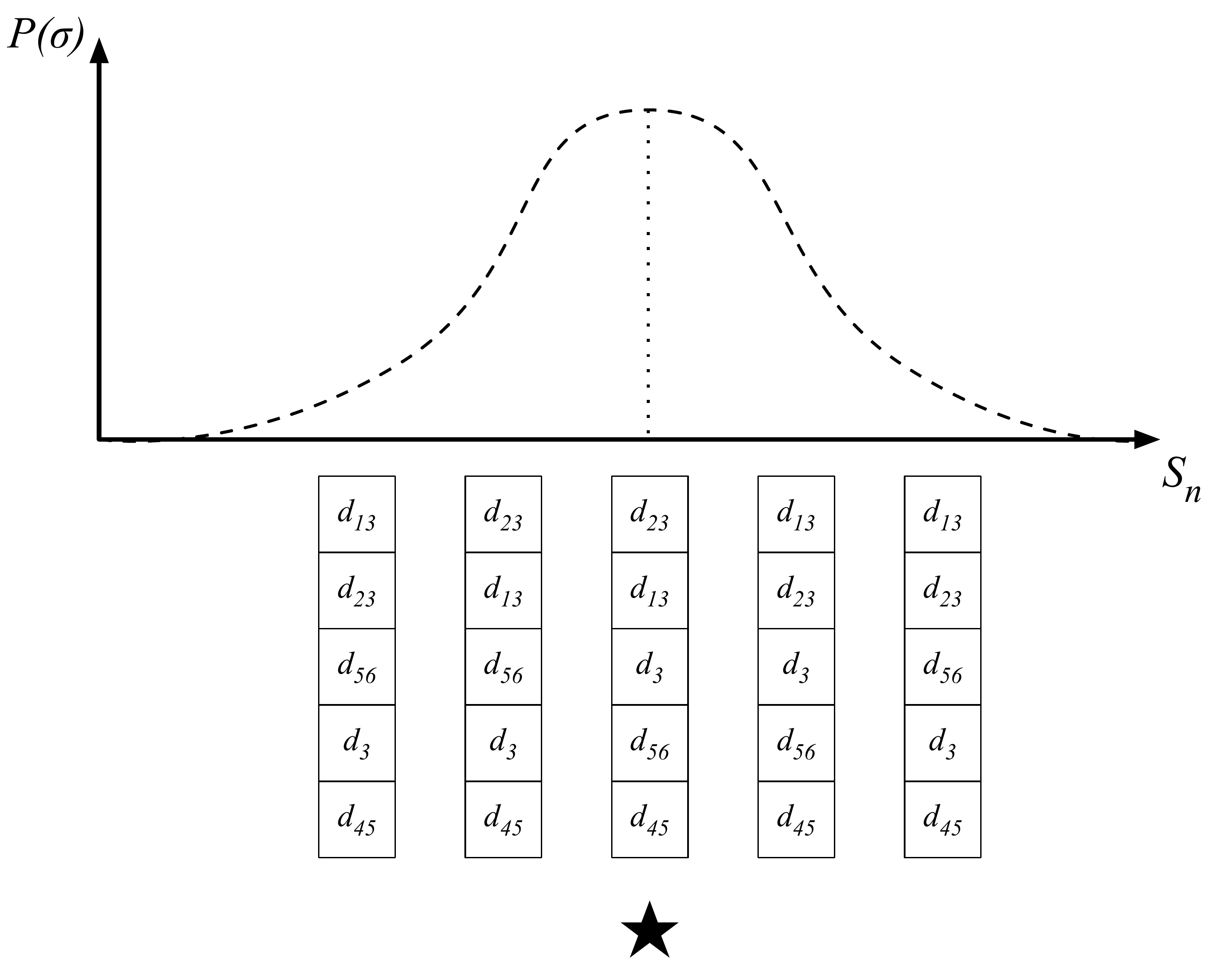}
\caption{Distribution over rankings.  Traditional evaluation methodologies consider only a single ranking (indicated by the $\star$) while stochastic rankers consider a distribution over rankings.}\label{fig:rank-dist-intro}
\end{figure}

Nonuniform exposure of relevant items resulting from ranking has multiple effects.
It strongly affects the allocation of user attention (and therefore content exposure, visibility, and consumption-related revenue) to results and their producers, giving rise to fairness concerns for content producers \cite{burke:multisided,asia:equity-of-attention,singh:exposure}; if there are qualitative differences in comparably-relevant results, systematically favoring results preferred by one group of users affects other groups' quality of service \cite{mehrotra:demographics-of-search} and may affect user retention \cite{liang:fairness-without-demographics}; similarly, in recall-oriented search or in scenarios where a searcher is interested in broad subtopical exposure, systematically promoting some relevant documents over others may risk overlooking important content; it can homogenize users' information experiences and promote rich-get-richer effects \cite{chaney:confounding}; and, although not often analysed in the design of algorithms, nonuniform exposure to relevant content may affect users' perception of the makeup of relevant information and its production community.

There may also be difference of degree: if there is a small difference in the relative relevance of two documents, but a large difference in the attention users tend to pay to the positions in which they are ranked, the ranking may amplify small difference in content value into a large difference in the producers' return for providing that value.

Unfortunately, providing a static ranking for a query (in retrieval) or context (in recommendation) limits the ability for an algorithm to distribute exposure amongst relevant items.
We propose to evaluate  information access systems using \emph{distributions over rankings} in response to a query or context. 
Figure \ref{fig:rank-dist-intro} depicts this approach.
More precisely, for a fixed query, we assume that a ranker, $\ranker$, samples a permutation $\permutation$ from a distribution over the set of all permutations $\permutationset$ of $\ndocs$ documents.  This allows it to provide equal exposure to relevant items \emph{in expectation}.
And while current evaluation metrics and methods measure the relevance or utility of a single ranking per query, with a distribution of rankings, we can compute the expected value of the metric.

This paper provides the foundation for an exposure-based approach to evaluating rankings and advocates for exploring the family of stochastic ranking policies within that framework.  
To that end, we 
\begin{inlinelist}
  \item define the concept of \emph{expected exposure} and ways to operationalize it;
  \item discuss its relationship to existing retrieval metrics, including diversity, novelty, and fairness metrics;
  \item apply it to measure item exposure under stochastic versions of existing retrieval and recommendation algorithms
\end{inlinelist}.
We argue that exposure provides a means of looking at several different concerns in the evaluation and impact of information access systems, and believe generalizing evaluation from deterministic rankers to stochastic rankers provides a broad area of study with implications for classic and contemporary problems in information access.

We begin by discussing the connection of previous work with our exposure-based evaluation and stochastic ranking (\S\ref{sec:relatedwork}).  We will then present the framework for evaluating with expected exposure and stochastic ranking together (\S\ref{sec:expected-exposure}).  These definitions of expected exposure have deep connections to existing metrics, which we describe in \S\ref{sec:exposure-metric-relationships}.  We then describe our experimental apparatus for analyzing these metrics in \S\ref{sec:analysis}.  We also propose a procedure to optimize towards these metrics in \S\ref{sec:optimization}.  We conclude with a discussion of our findings (\S\ref{sec:discussion}).  
\section{Related Work}
\label{sec:relatedwork}
Our work is inspired by and draws together two areas of work:
\begin{inlinelist}
\item metrics recently developed in the context of algorithmic fairness, and
\item randomized ranking algorithms developed in the context of online learning and optimization.
\end{inlinelist}

\subsection{Fairness}

Exposure optimization has been proposed as a means of achieving fairness in ranking: fairness for \emph{individuals} means that exposure should be proportional to relevance for every subject in a system~\cite{asia:equity-of-attention}, while fairness for \emph{groups} means that exposure should be equally distributed between members of groups defined by sensitive attributes such as gender or race~\cite{singh:exposure}.
From an optimization point of view, \citet{singh2019policy} and \citet{yadav2019fair} consider a similar notion of exposure fairness over multiple rankings as our work.  
Our work situates exposure-based measures in the context of information retrieval evaluation, allowing us to 
\begin{inlinelist}
  \item extend them with user models from existing retrieval metrics, 
  \item relate them with the objectives and formalisms of other retrieval metrics, and
  \item introduce a new experimentation protocols based on stochastic ranking.
\end{inlinelist}

\citet{gao:fair-ranking} recently proposed a randomized policy for diversifying search results very similar to our work, albeit in the context of group fairness.  While studying connection between fairness and diversity empirically, we attempt to more formally elucidate the relationship and study broader connections beyond group fairness.  

Beyond the definitions explicitly focusing on exposure, other fairness definitions in practice lead to enhanced equality of exposure, for instance, by requiring equal proportions of individuals from different groups in ranking prefixes~\cite{celis:ranking-w-fairness-constraints, zehlike:fa-star-ir}.  Similarly, \citet{yang:fair-ranking} measure fairness by computing sum of position-discounted set-wise parity at different rank thresholds.  \citet{beutel:pairwise-fairness} approach fair ranking by conducting pairwise analysis of user engagement with the protected groups in a ranking.  \citet{zehlike:fair-ltr} propose a supervised learning to rank method to optimize for fair exposure but focus only on the top position in ranking.  It is not obvious how their proposed approach can be extended beyond the first rank position.  In constrast to this literature, we study metrics that have clear user behavior model amenable to extension.

The notion of \emph{meritocratic fairness}, originally introduced as a fairness objective for online bandit learning~\cite{joseph:fair-bandits} and then applied to the problem of selecting a group of individuals from incomparable populations~\cite{kearns:meritocratic-fairness}, intuitively requires that less qualified candidates do not have a higher chance of getting selected than more qualified candidates.
In our setting, this translates to ensuring that less-relevant documents are not likely to be ranked above more-relevant documents.
Our construct of target exposure connects this work to meritocratic fairness, in that a system satisfying equity of expected exposure will satisfy the goals of meritocratic fairness by allocating more exposure to relevant documents than to non-relevant documents, it also imposes a stronger constraint by requiring documents with comparable relevance to receive comparable exposure, preventing runaway popularity feedback loops that meritocratic fairness allows.

\subsection{Stochastic Ranking}
Randomization (either explicit or implicit) is ubiquitous in many information access systems and has been shown to be useful for eliciting user feedback and lead to desirable system properties.
\citet{pandey:random-rankings} first proposed randomized ranking motivated by click exploration.
Further strategies \cite{radlinski:active-exploration, radlinski:minimally-invasive-randomization, hofman:thesis, wang2016learning} have been developed following this approach for collecting unbiased feedback for learning to rank.
Instead of using randomization to collect unbiased training data, \citet{joachims2017unbiased} use it to estimate the parameters of a click propensity model that allows ranking models to be trained on biased feedback.
Using randomness in ranking may also be a means of improving diversity \citep{radlinski:icml08}.

Recently, \citet{bruch:stochastic-ltr} demonstrate that learning to rank models can be optimized towards expected values of relevance metrics computed over multiple rankings sampled based on estimated relevance.  While not developed in the context of deploying a stochastic ranker, we adopt some of the methodologies therein in our experiments.

\section{Expected Exposure}
\label{sec:expected-exposure}
Given a query, we are interested in measuring the expected exposure of an item to a searcher with respect to items of similar relevance.  
Specifically, we would like to define a metric that quantifies a system's deviation from an ideal expected exposure of items of the same relevance.  
To this end, we adopt the following principle of \emph{equal expected exposure},\footnote{This principle is related to \emph{equity of attention} \cite{asia:equity-of-attention}, which also ties exposure to relevance. However, \emph{equity of attention} was originally amortized across information needs. While this paradigm accounts for changing relevance, the system might increase exposure of items for inappropriate information needs. Thus, in this paper we propose to measure exposure per information need. In this sense, the distinction between equal expected exposure and equity of attention is similar to the difference between macroaveraging and microaveraging of relevance metrics.}
\begin{quote}
\emph{Given a fixed information need, no item should be exposed (in expectation) more or less than any other item of the same relevance.}
\end{quote}
This principle complements the existing core principle of ranked retrieval that more relevant documents should appear before less relevant documents.  
In this section, we will introduce an evaluation methodology based on the principle of equal expected exposure.  

We note that existing relevance metrics do not measure the extent to which systems satisfy this principle, as they typically ignore differences in exposure amongst items of the same relevance.  
As a result, existing relevance metrics will not be able to distinguish a system that satisfies this principle from one that does not.

We will start by remaining agnostic about how items are exposed to searchers, only that there is some way in which searchers interact with a ranking of items that is related to the exposure. 
More formally, let a ranking be defined as a permutation of the $\ndocs$ documents in the corpus.  
The set of all permutations of size $\ndocs$ is referred to as the symmetric group or $\permutationset$ in abstract algebra.  
Given a query $\query \in \queries$ with $\nreldocs$ relevant documents, an optimal permutation would place some ordering of the $\nreldocs$ relevant items at the top positions, followed by some ordering of the $\nnreldocs$ nonrelevant documents.  
Per existing models, exposure monotonically---often-exponentially---decreases with position in a ranking.  
Therefore, for a static ranking, we can see that 
\begin{inlinelist}
	\item some relevant documents receive more exposure than other relevant documents, and
	\item some nonrelevant documents receive more exposure than other nonrelevant documents.
\end{inlinelist}
A static ranking will therefore always violate equal expected exposure.  
Unfortunately, classic retrieval systems only provide and are evaluated according to static rankings.  

However, we know that there are $\nreldocs!\nnreldocs!$ optimal rankings.  
If an oracle provided us with an optimal ranking at random, any relevant item would be ranked in position $0\leq i <\nreldocs$ with the same probability.\footnote{Note that we use base-0 ranks throughout this manuscript.}  
As a result, all relevant items would receive the same exposure in expectation; similarly all nonrelevant items would receive the same exposure in expectation.  
Such a oracle would satisfy equal expected exposure.  
We will refer to the expected exposure of all items under the oracle policy as the \emph{target exposure}, represented as a $\ndocs\times 1$ vector $\targetexposure$.

Just as we can satisfy ideal expected exposure by using a stochastic oracle, a retrieval system can improve the distribution of exposure by using a \emph{stochastic policy}, a protocol where, in response to a query, a distribution over rankings is provided.  
Formally, given a query $\query$, a ranking policy $\ranker$ provides a distribution over all permutations, $\sum_{\permutation\in\permutationset}\ranker(\permutation | \query) = 1$.  
Classic ranking algorithms are a special case which only assign probability to a single, static permutation.  
We will refer to such an algorithm as a \emph{deterministic policy}.
We note that most classic evaluation metrics (e.g. mean average precision) only evaluate a single, static permutation from a deterministic policy.

Given a policy $\ranker$ and a model of how the searcher might interact with a ranking, we can compute the expected exposure of all of the items in the corpus.  We will represent the expected exposure of all items under $\ranker$ as a $\ndocs\times 1$ vector $\exposure$.

In order to measure the deviation from equal expected exposure, we compare the target exposure $\targetexposure$ and sytem exposure $\exposure$.  
One simple way to do this is to compute the squared error between  $\targetexposure$ and $\exposure$,
\begin{align}
	\eeLoss(\exposure, \targetexposure) &= \|\exposure - \targetexposure\|_2^2 \label{eq:ee}\\
	 &= \underbrace{\|\exposure\|^2_2}_{\eeDisparity} - \underbrace{2\fdtrans{\exposure}\targetexposure}_{\eeRelevance}+ \|\targetexposure\|^2_2 
\end{align}
where $\eeDisparity$ or \emph{expected exposure disparity} measures inequity in the distribution of exposure; $\eeRelevance$ or \emph{expected exposure relevance} measures how much of the exposure is on relevant documents;  the remaining term is constant for a fixed information need.  

This derivation allows us to clearly decompose expected exposure into a relevance and disparity components.   A system that achieves optimal $\eeRelevance$ may maximize disparity (e.g. a static ranking with all relevant items at the top).  Similarity a system that minimizes $\eeDisparity$ will have very bad expected exposure relevance (e.g. a random shuffling of the corpus every time a query is submitted).  

We empirically observed (in \S\ref{sec:analysis-results}) a tradeoff between the disparity ($\eeDisparity$) and relevance ($\eeRelevance$).  This tradeoff is often controllable by a parameter in a stochastic policy that affects the degree of randomization.  So, at one extreme, the parameter results in a  deterministic policy that can achieve high relevance but also incurs high disparity.  At the other extreme, the parameter results in a policy that randomly samples from amongst all permutations, achieving the lowest disparity but the lowest relevance.  Given that such a parameter can often be swept between a minimum and maximum disparity, we can plot a disparity-relevance curve reflecting the nature of this tradeoff.  We use the area under this curve, $\eeAUC$, as a summary statistic of this curve.  

While we expect $\eeRelevance$ to behave similar to traditional relevance-based metrics--especially those sharing similar assumptions about how searchers interact with a ranking, reasoning about relevance and disparity within a single formalism allows us to compose aggregate metrics like $\eeAUC$, which traditional metrics do not capture (\S\ref{sec:analysis-results}).

\subsection{Computing Exposure with User Browsing Models}
So far, we have remained agnostic about how items are exposed to users.  
In this section, we will describe how we can compute the exposure vector $\exposure$ for an arbitrary ranker, including the oracle ranker.  
Unlike previous fair ranking metrics, we approach exposure by adopting user models from existing information retrieval metrics.  
We focus on models from two metrics, rank-biased precision and expected reciprocal rank, although this analysis can be extended to more elaborate browsing models \cite{fdiaz:robust-mouse-tracking-models}.

\emph{Rank-biased precision (RBP)} is a metric that assumes that a user's probability of visiting a position decreases exponentially with rank \cite{moffat:rbp},
\begin{align}
	\rbp(\permutation) &= (1-\rbpp)\sum_{i\in [0,\rbpd)} \qrels_{\permutation_i} \rbpp^{i}
\end{align}
where $\qrels$ is the $\ndocs\times 1$ binary relevance vector; $\rbpp$ is referred to as the \emph{patience parameter} and controls how deep in the ranking the user is likely browse; and $\rbpd$ is the maximum browsing depth.  The multiplicative factor $1-\rbpp$ ensures that the measure lies in the unit range.

We consider that the expected exposure of a document $\doc$ is computed, in expectation, as,
\begin{align}
	\exposure_{\doc}&=\sum_{\permutation\in\permutationset} \ranker(\permutation|\query) \rbpp^{\invpermutation_\doc}
\end{align}
where $\invpermutation$ is a map from document indexes to ranks.  
This allows us to compute $\exposure$ for an arbitrary policy $\ranker$.  

Recall that the oracle policy selects randomly amongst all rankings with all of the relevant documents at the top.  
Since each document occurs at each of the top $\nrel$ positions equally, the target expected exposure for a relevant document is,
\begin{align*}
	\targetexposure_{\doc}&=\frac{1}{\nrel}\sum_{i\in[0,\nrel)}\rbpp^{i}\\
	&=\frac{1-\rbpp^{\nrel}}{\nrel(1-\rbpp)}\nonumber
\end{align*}
Since the set of nonrelevant documents is usually very large, all nonrelevant documents will have equal expected exposure close to zero.

\emph{Expected reciprocal rank (ERR)} is a metric that assumes that a user's probability of visiting a position is dependent on how many relevant documents appear a earlier positions \cite{chapelle:err}.  
The intuition is that earlier relevant documents may satisfy the user and prompt them to stop scanning the ranking.  
We adopt generalized expected reciprocal rank, a model which incorporates a patience parameter similar to that used in RBP \cite[\S7.2]{chapelle:err}.  
\begin{align}
	\err(\permutation) &= \sum_{i\in[0,\errd)} \errphi(\qrels_{\permutation_i})\prod_{j\in[0,i)}\rbpp(1-\errphi(\qrels_{\permutation_j}))
\end{align}
where $\errphi$ converts relevance to a probability of stopping the browsing.  Normally this is zero for nonrelevant documents and some value between 0 and 1 for relevant documents.  
As with RBP, the expected exposure of document $\doc$ can be computed as,
\begin{align*}
	\exposure_{\doc}&=\sum_{\permutation\in\permutationset} \ranker(\permutation|\query) \errp^{\invpermutation_\doc}\prod_{j\in[0,\invpermutation_\doc)}(1-\errphi(\qrels_{\permutation_j}))
\end{align*}

Similarly, the target expected exposure of a relevant document is,
\begin{align*}
	\targetexposure_{\doc}&=\frac{1}{\nrel}\sum_{i\in[0,\nrel)}\errp^{i}(1-\errphi(\qrels_{\rdoc}))^{i}\\
	&=\frac{1-\errp^{\nrel}(1-\errphi(\qrels_{\rdoc}))^{\nrel}}{\nrel(1-\errp(1-\errphi(\qrels_{\rdoc})))}\nonumber
\end{align*}
and close to zero for nonrelevant documents.

\subsection{Extension to Graded Judgments}
\label{sec:graded}
So far, we have focused on binary relevance.  For graded judgments, the ideal ranker always orders documents correctly by grade.  We take all permutations satisfying this requirement and assume the ideal ranker has nonzero support only for these values.  We then compute the expected exposure for documents by grade.  Let $\nrel_{\grade}$ be the number of documents with relevance grade $\grade$ and $\nrel_{>\grade}$ the number of documents with relevant grade strictly larger than $\grade$.  Without loss of generality, assume that grades take integer values.  Given an RBP browsing model, the optimal exposure for a document $\doc$ with grade $\grade$ is,
\begin{align*}
	\targetexposure_{\doc}&=\frac{1}{\nrel_{\grade}}\sum_{i\in[\nrel_{>\grade},\nrel_{>\grade-1})}\rbpp^{i}\\
	&= \frac{\rbpp^{\nrel_{>\grade}}-\rbpp^{\nrel_{>\grade-1}}}{\nrel_{\grade}(1-\rbpp)}
\end{align*}
The derivation for the ERR user model is similar.

We note that this extension assumes that the a searcher will always prefer to see items with higher grade.  In situations where, for example, the grade of an item is inversely correlated with some important property of a document (e.g. a subtopic, authors from underrepresented groups), then these groups will be under-exposed.  In such cases, an alternative definition of $\targetexposure$ may be more appropriate (see \S\ref{sec:fairness}).

\section{Relationship to Other Metrics}
\label{sec:exposure-metric-relationships}
Expected exposure, both in motivation and in definition, has connections to existing retrieval metrics.  In this section, we will discuss those relationships, highlighting the unique properties that expected exposure measures.  

\subsection{Retrieval Metrics}
Measures such as RBP and ERR could be considered {\em precision metrics}, as they reward rankers for retrieving relevant material higher in the ranking.
While based on the same user model, it is {\em not} the case that optimizing $\rbp$ will also minimize Equation \ref{eq:ee}, even if exposure is based on an RBP browsing model.  
To see why, consider a deterministic policy that outputs a static optimal ranking.  
Although $\eeRelevance$ will be optimal, $\eeDisparity$ will be very large since exposure is concentrated at the top ranks.  Indeed, the value of $\eeDisparity$ for a static optimal ranking will be as bad as a static ranking that places all of the relevant document at the bottom since disparity is based only on the exposure and not on relevance.  
The converse, that minimizing Equation \ref{eq:ee} also optimizes $\rbp$, {\em is} true.  If expected exposure is based on the RBP user model, a system that optimizes expected exposure will essentially be shuffling relevant documents at the top of the ranking and nonrelevant items in the bottom, just as with the oracle in \S\ref{sec:expected-exposure}.

Optimizing recall means focusing on ensuring that all of the relevant items in the corpus occur at high ranks.
Several of our motivating examples might be considered addressable by a retrieval system optimized for high recall (e.g.\ e-discovery, academic search, systematic review).  However, if we assume, as many user models do, that a user may terminate their scan of a ranking early, then there is a chance that even a high-recall system, especially in situation where there are numerous relevant documents, a user will not be exposed to all relevant items.  As a result, we would argue that expected exposure reduces the risk of overlooking a relevant document.

\subsection{Fairness}
\label{sec:fairness}
Algorithmic fairness, in the context of information retrieval and recommendation,  deals with the treatment of individuals associated with retrievable items \cite{burke:multisided}.  These might be document authors in text retrieval, job candidates in recruiting, or musicians in song recommendation.

\textbf{Individual Fairness.} 
Expected exposure is closely related to various notions of individual fairness that quantify the extent to which models are fair to all individuals. Dwork et al. defined individual fairness in the context of classification models seen as mappings from individuals to probability distributions over outcomes~\cite{dwork:individual-fairness}. In this setting, individual fairness is defined using the Lipschitz condition: the distributions of classification outcomes $P_C$ of two  individuals $u_1, u_2$ who are sufficiently similar according to a chosen similarity metric $d$ should be close according to a distribution similarity metric $D$. Formally, if $d(u_1, u_2) < \delta$, then $D(P_C(u_1), P_C(u2)) < \Delta$.
When will a retrieval policy be individually fair according to this definition?
Assume we define $\delta$ and $d$ such that two documents of equal relevance grade satisfy the above inequality, and two documents of different relevance grades do not.
Assume furthermore that outcomes are measured as the expected exposure of individual documents. A stochastic ranker that distributes exposure (almost) equally among the documents of equal relevance grades (in particular if it achieves optimal expected exposure according to Eq.~\ref{eq:ee}) is individually fair according to the above definition. However, the reverse does not hold: It is possible that an individually fair and an unfair stochastic rankers lead to similar values of the expected exposure measure (the total loss value in Eq.~\ref{eq:ee} can be aggregated equitably from documents of the same relevance level or from only few documents within a relevance grade).

\textbf{Group Fairness.} We can use exposure to define a group notion of provider fairness  by measuring whether deviation from expected exposure differs between different groups of documents (or their authors). Let $\intents$ be the set of $\nintents$ attributes that a document might be associated with.  Attributes may be related to, for example,  demographic or other group information about the provider.  Let $\attributeMatrix$ be a $\ndocs\times\nintents$ binary matrix of the group identity associated with each document in the corpus.  We can then compute the total exposure for all documents with an attribute by $\iexposure_\attributeMatrix = \fdtrans{\attributeMatrix}\exposure$.  If we are interested in equal exposure across groups, we can define the target group exposure as $\targetiexposure_{\fde} = \exposureMassConstant\fdtrans{\attributeMatrix}\fde$ where $\fde$ is a $\nintents\times 1$ vector of ones and $\exposureMassConstant$ is a normalizing constant based on the total exposure given a browsing model.  We can then use Equation \ref{eq:ee} as a measure of \emph{demographic parity}  \cite[\S4.1]{singh:exposure}.  If desired, we can replace $\fde$ with some other distributions, such as population level proportions \citep{piotr:fair-rank}.   Target exposures like $\targetiexposure_{\fde}$ only balance group representation, but some groups may produce more relevant content than others.  If we are interested in exposure proportional to relevance, we can define the target exposure as $\targetiexposure=\fdtrans{\attributeMatrix}\qrels$, referred to as \emph{disparate treatment} \cite[\S4.2]{singh:exposure}.  Finally, if we are interested in ensuring the exposed items are relevant, we can define a new matrix $\intentMatrix = \text{diag}(\qrels)\attributeMatrix$ and exposure vector $\iexposure_\intentMatrix = \fdtrans{\intentMatrix}\exposure$.  If we let $\targetiexposure_\intentMatrix=\exposureMassConstant\fdtrans{\intentMatrix}\qrels$, then we recover \emph{disparate impact}  \cite[\S4.3]{singh:exposure}.

\subsection{Topical Diversity}
Exposure metrics are closely related to topical diversity metrics \cite{santos:diversity-survey}.  One common way to measure topical diversity is to consider so-called `intent-aware' metrics defined  as,
\begin{align*}
	\ia(\permutation)	&= \sum_{\intent\in\intents} p(\intent|\query) \metric(\permutation|\intent)
\end{align*}
where $\metric(\permutation|\intent)$ computes a standard metric considering only those documents with aspect $\intent$ as relevant.  The intent-aware RBP metric is defined as
\begin{align*}
	\iaRBP(\permutation) &= \sum_{\intent\in\intents} p(\intent|\query) (1-\rbpp)\sum_{i\in [0,\rbpd)} \iqrels_{\permutation_i} \rbpp^{i}\\
\end{align*}
If we assume that $p(\intents|\query)$ is proportional to the frequency of $\intent$ in the set of relevant documents, then $\iaRBP(\permutation)\propto\fdtrans{\iexposure}_{\intentMatrix}\targetiexposure_{\intentMatrix}$.   In other words, topic diversity reduces to a scaled  relevance term in the disparate impact metric (\S\ref{sec:fairness}).  In the event that we are interested in uniform $p(\intent | \query)$, then we can redefine the target exposure accordingly and recover the relevance term in the demographic parity metric.   Both of these formulations ignore $\eeDisparity$ and it is worth observing that intent-aware metrics often include a `subtopic recall' factor to \cite{sakai:d-sharp-11} to ensure that all subtopics are retrieved.  We believe that the disparity term captures precisely this behavior.

\section{Metric analysis}
\label{sec:analysis}
We are interested in empirically studying the $\eeDisparity$ and $\eeRelevance$.  Specifically, we will answering the following questions in our experiments: 
\begin{inlinelist}
  \item can the metric distinguish between different randomization strategies?
  \item does an exposure-based relevance metric measure something different from a static ranking metric based on the same user model?

\end{inlinelist}

\subsection{Randomizing a Deterministic Policy}
\label{sec:analysis-model}
The focus of this paper is on evaluation.  However, we were interested in studying our metrics for stochastic rankers, which are not readily available outside of specialized online learning environments.  As such, we developed several stochastic rankers for our experiments based on post-processing a precomputed set of retrieval scores.

\paragraph{Plackett-Luce (PL)}
Our first randomization strategy uses Plackett-Luce sampling to sample a permutation \cite{luce:choice-axiom,plackett:analysis-of-permutations}.  To do this, we create a multinomial distribution $p(\doc|\query)$ over the corpus using the $\ell_1$ normalization of the retrieval scores.  The Plackett-Luce model samples a permutation by first sampling the document at position 0 using $p(\doc|\query)$.  We then set the probability of the selected document to 0, renormalize, and sample the document at position 1 from this modified distribution.  We continue this process until we exhaust the scored documents.  In order to control the randomness of the process, we use a modified sampling distribution,
\begin{align*}
	\prob(\doc|\query) &= \frac{\score^{\alpha}_{\doc}}{\sum_{\doc'}\score^{\alpha}_{\doc'}}
\end{align*}
where $\alpha\geq 0$.  When $\alpha=0$, all permutations are equally likely and $\eeDisparity$ is minimized; 
as $\alpha$ increases $\ranker$ concentrates around original static ranking and disparity degrades.  We refer to this as the Plackett-Luce (PL) policy.

\paragraph{Rank Transpositions (RT)}
Our second randomization strategy ignores the retrieval scores and samples permutations by shuffling the original ranked list.  We shuffle by repeatedly sampling pairs of positions and swapping the documents.  Such a process takes $\frac{1}{2}\ndocs\log\ndocs + c\ndocs$ iterations to converge to sampling a random permutation \cite{diaconis:random-transpositions-mixing-time}.  This is precisely a random walk on $\permutationset$ where permutations are connected by pairwise transpositions.   As such, we can introduce a `restart probability' to teleport the random walker back to the original ranked list.  If this probability is $\rwRestartProbability$, then the number of steps of the random walk follows a geometric distribution with support $[0,\infty)$.  Our randomization strategy then first samples the number of steps $\rwNumSteps$ from the geometric distribution and then conducts $\rwNumSteps$ random transpositions.  We refer to this as the rank transposition (RT) policy.

These two methods are intentionally constructed to perform differently.  The PL policy takes a deterministic policy's scores into consideration and will, therefore, be more conservative in removing high-scoring items from the top of the ranked list.  The RT policy, on the other hand, randomly swaps pairs, regardless of score or position.  As a result, we suspect that the PL policy should outperform the RT policy, given a fixed base deterministic policy.

\subsection{Method}
\label{sec:analysis-method}
We analyze the behavior of expected exposure metrics using the postprocessing of deterministic policies in two domains.  The first is based on archival TREC submissions focus in information retrieval conditions.  The \robust dataset consists of 440 runs submitted to the TREC 2004 Robust track which evaluated systems on a set of 249 queries and binary relevance labels.  We adopt this dataset because it has been well-studied in the context of evaluation metrics.  

Our second dataset, \ml, is a movie recommendation dataset consisting of 25 million ratings of 59 thousand movies by 163 thousand users~\cite{harper:movielens}.  We used LensKit 0.8.4 \cite{ekstrand:lkpy} to generate runs representing binary implicit-feedback matrix factorization (IMF) \cite{pilaszy:als} and Bayesian personalized ranking (BPR) \cite{rendle:bpr}.\footnote{BPR is implemented by the \texttt{implicit} package (\url{https://github.com/benfred/implicit}).} We adopt implicit feedback instead of ratings in order to study the behavior of expected exposure under binary relevance.

We use a $\rbpp=0.50$ for all of our experiments, as consistent with standard TREC evaluation protocol.  RBP and ERR are evaluated at depth 20.   For stochastic rankers, we sample 50 rankings during evaluation to estimate expected exposure metrics.  We found that this was sufficient to converge to appropriate expected metric values.  Experiments randomizing deterministic policies rerank the top 100 documents from the original static ranking.   

\subsection{Results}
\label{sec:analysis-results}
\begin{figure}[t]
    \centering
\includegraphics[width=2.85in]{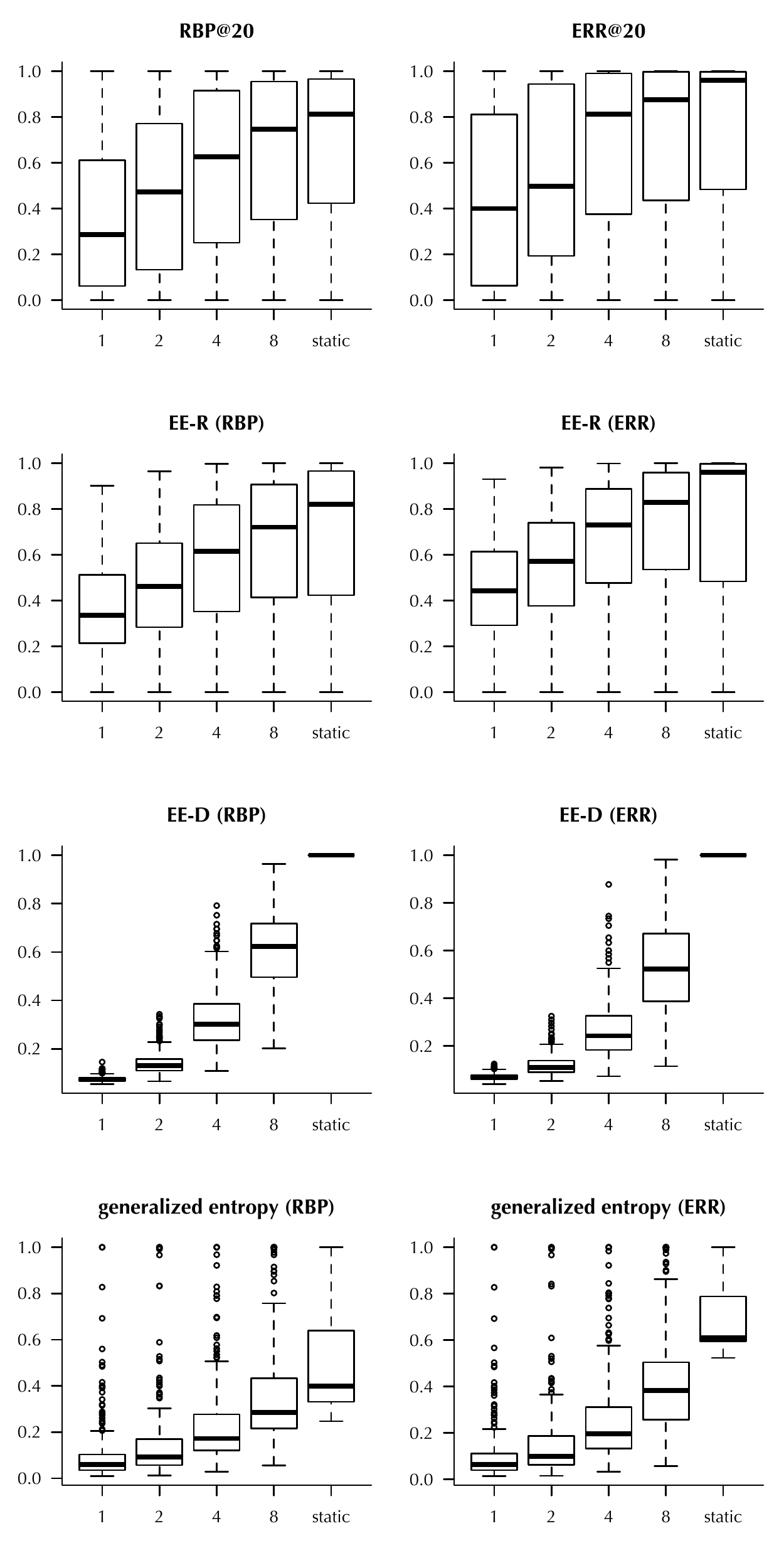}\\
    \caption{Behavior of expected exposure metrics for a deterministic run from the \robust dataset randomized using the Plackett-Luce model.      
    Each horizontal axis indicates the value of $\alpha$, where lower values indicate more randomization.  
    Each vertical axis reflects the performance of policies on static ranking relevance metrics (top row), expected exposure relevance metrics (second row), expected exposure disparity metrics (third row), and generalized entropy on the relevant set of documents (fourth row) using two browsing models (left: RBP; right: ERR).  }    \label{fig:examples}
\end{figure}

Before analyzing our metrics in aggregate, we present our metrics on an example run from \robust.  In Figure \ref{fig:examples}, we show the behavior of our randomization model for $\eeRelevance$ and $\eeDisparity$, under both the ERR and RBP user models.  We compare these metrics to $\rbp$ and $\err$, two classic static ranking metrics.  We also measure the generalized entropy of exposure on the relevant set of documents \cite{speicher:gh-fairness}; this allows us to assess the disparity amongst relevant items.  

Comparing classic metrics and $\eeRelevance$ in the first and second rows, we observe correlated behavior as randomization changes.  Across a sample of runs, we found that the expected $\rbp$ and $\eeRelevance$ were strongly correlated ($r=0.99$, $p<0.01$); a perfect correlation was observed between expected $\err$ and $\eeRelevance$ with an ERR model.  This is unsurprising given that the relevance factor in the expected exposure metric is precisely the expectation of the static ranking metric.  The imperfect correlation for RBP is due to normalization term in the classic RBP model.

Comparing generalized entropy and $\eeDisparity$, we also observe correlated behavior across both RBP ($r=0.47$, $p<0.01$) and ERR user models ($r=0.65$, $p<0.01$).   Because the generalized entropy is computed over only relevant documents,  this suggests that  $\eeDisparity$ is sensitive to changes in expected exposure to relevant documents, not the dominant, nonrelevant set.

\begin{figure}[t]
    \centering
\includegraphics[width=1.6in]{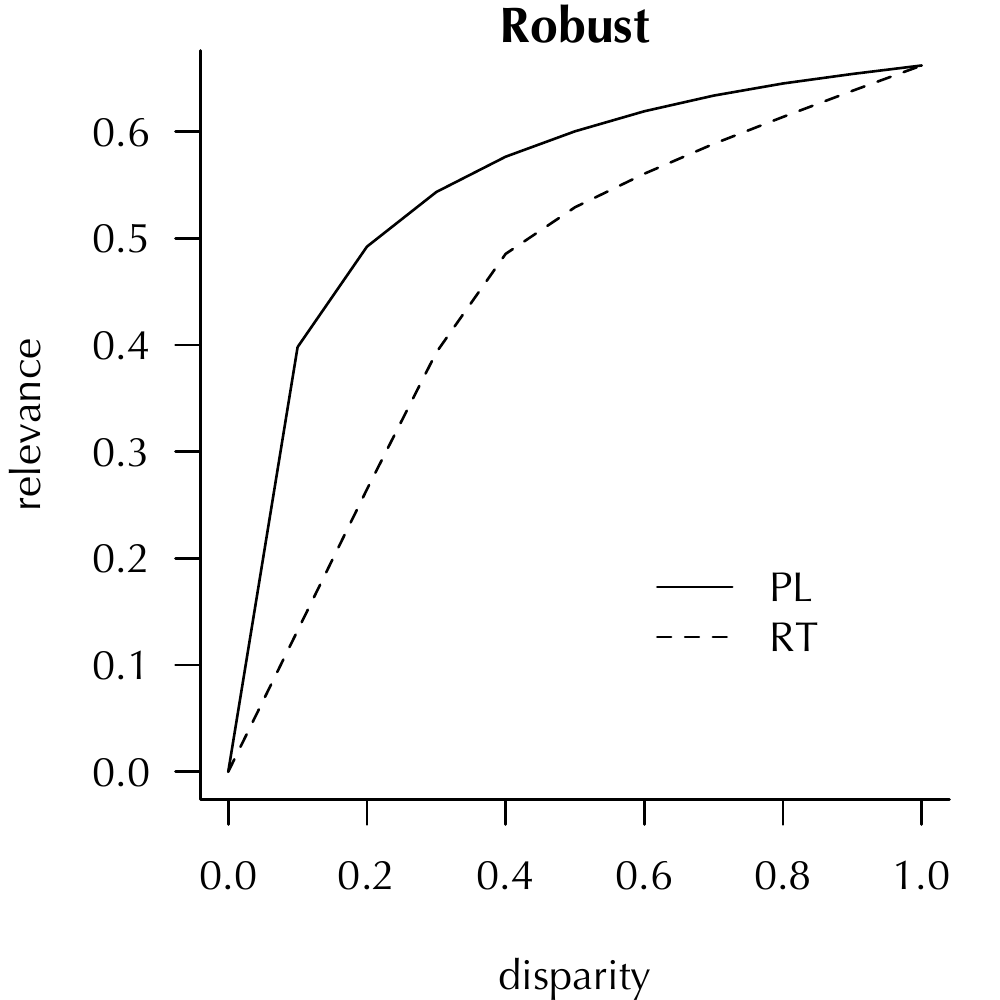}\includegraphics[width=1.6in]{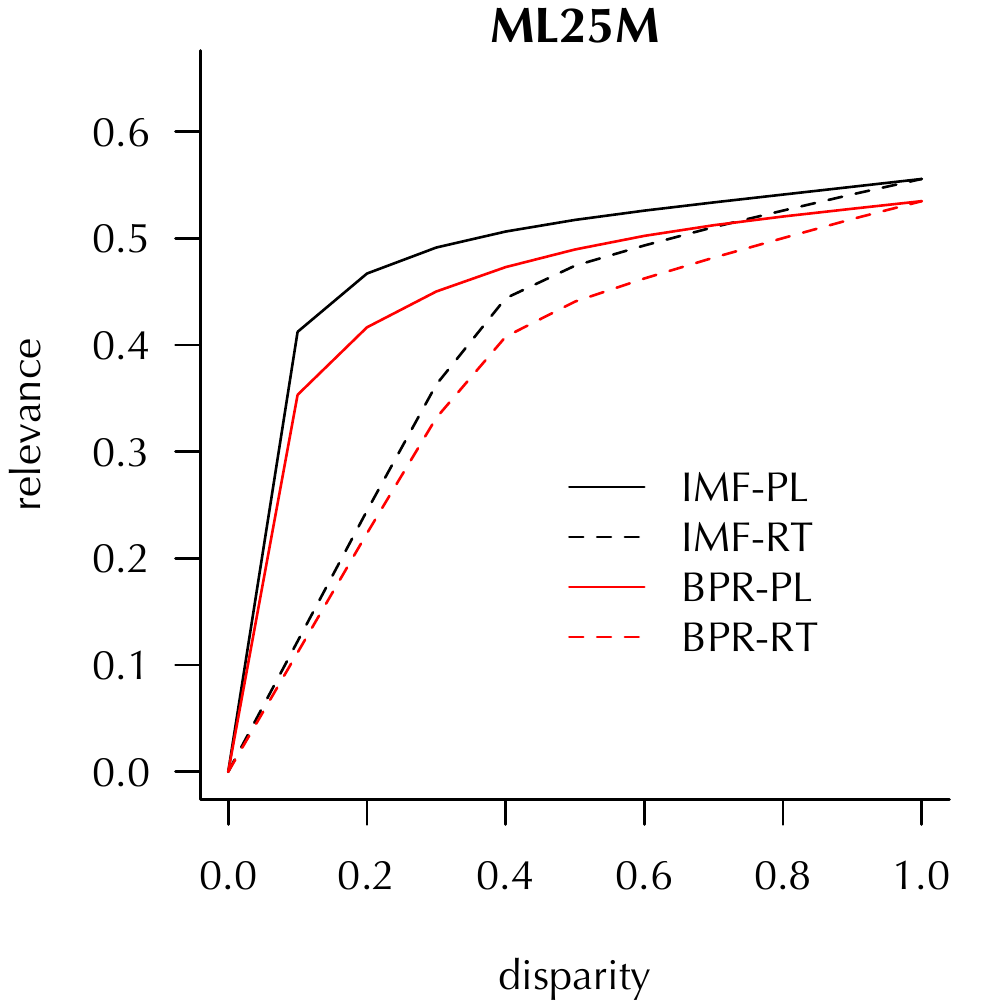}\\
    \caption{Disparity-relevance tradeoff curve for a random \robust run and our two recommendation runs on \ml with Placket-Luce randomization and rank transposition randomization.  }
    \label{fig:example-dr}
\end{figure}

Comparing the behavior of $\eeDisparity$ and $\eeRelevance$ in Figure \ref{fig:examples}, we notice the disparity-relevance tradeoff mentioned in \S\ref{sec:expected-exposure}.  In order to visualize this tradeoff more clearly, we present  example disparity-relevance curves for randomization of an arbitrary \robust run and our two recommender systems on \ml in  Figure \ref{fig:example-dr}.  Disparity-relevance curves, when randomizing the same base policy, will have the same value of $\eeRelevance$ for $\eeDisparity=1$ because this recovers the original static ranking.  Similarly, all disparity-relevance curves begin with $\eeRelevance=0$ at $\eeDisparity=0$ because a completely random ranker will achieve minimal relevance by dint of the number of nonrelevant documents in the corpus (i.e. a random shuffle will mean that, in expectation, every document receives a tiny amount of attention).  Turning to the randomization policies being studied, across both domains and multiple runs, PL randomization policies dominate RT policies across all disparity points, confirming our intuition that incorporating score information improves post-processing performance.  This provides us with the ability to test the ability of exposure to distinguish between stochastic policies. 

\begin{figure}[t]
    \centering
\includegraphics[width=2.5in]{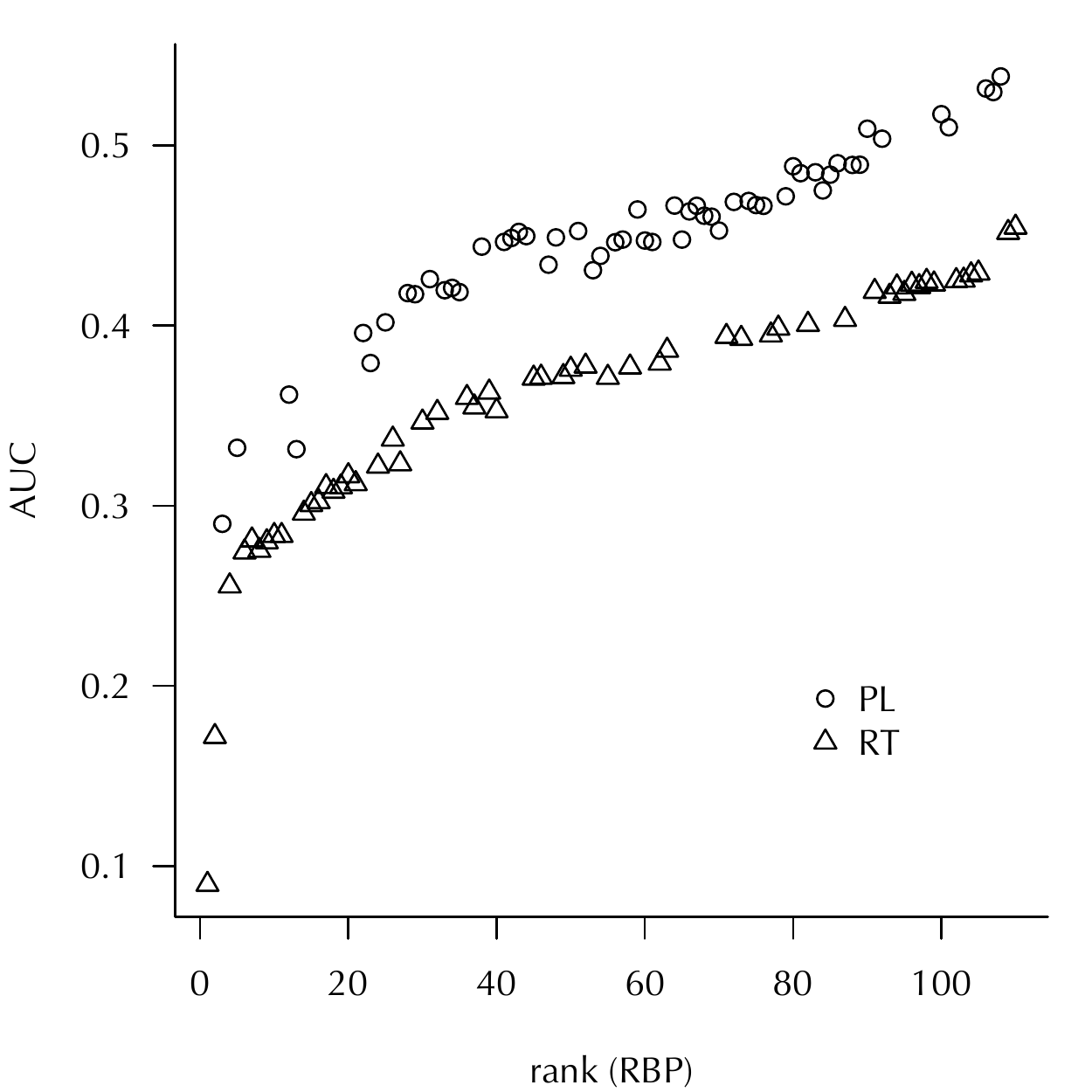}\\
    \caption{Sorting PL and RT runs by RBP.  Half of the runs submitted to \robust were subjected to PL randomization and half to RT randomization.  Runs were ranked by the RBP of the original, static ranking.  EE-AUC for the randomized runs, according to its treatment, is plotted on the vertical axis.}
    \label{fig:auc-vs-rbp}
\end{figure}

Given two stochastic rankers, we are interested in understanding whether our exposure metrics can more accurately identify  the superior algorithm compared to a metric based on a static ranking.  To that end, we randomly assigned the runs for the \robust dataset to either PL or RT randomization.  This provided us with $\eeAUC$ for each run as well as an RBP value for its base deterministic policy.  We ordered runs by the RBP and then inspected the $\eeAUC$ for the associated run.  In Figure \ref{fig:auc-vs-rbp}, we can see that, while RBP, a metric based on a static ranking, can approximately order runs for a fixed randomization policy ($\tau=0.89$, $p<0.05$), it cannot distinguish between the PL and RT policies.

\section{Optimizing for Expected Exposure}
\label{sec:optimization}

In the previous section, we introduced post-processing techniques to build stochastic rankers.  Given a model that is perfectly able to predict relevance, Plackett-Luce randomization should perform optimally, especially for binary relevance.  As such, a classic pointwise learning to rank  model \citep{cossock2006subset} with Plackett-Luce randomization may be an effective approach for expected exposure.  Moreover, calibration of relevance does not happen with pairwise learning to rank models  \citep{burges2005learning} and so we would expect these models, even if perfect, to perform worse than pointwise models, even with Plackett-Luce randomization.  However, learning to rank models are not perfect estimators of relevance.  Therefore, we believe there should be some advantage to optimizing directly for expected exposure.

In this section, we will examine the relationship between the performance of these approaches in the context of graded relevance as well as demographic parity (\S\ref{sec:fairness}).  We focused on a shared model architecture with varying loss functions in order to measure differences due to the objective alone, instead of artifacts resulting from the functional form of the models.  We begin by describing how we optimize for expected exposure before proceeding to our empirical results.

\subsection{Algorithm}
Although optimizing for pointwise or pairwise loss has been well-studied in the information retrieval community, directly optimizing for a metric based on a distribution over rankings has received less attention.  

We begin by defining an appropriate loss function for our model.  Turning to Equation \ref{eq:ee}, we can drop the constant term and add a hyperparameter to balance between disparity and relevance,
\begin{align}
    \instancelossmod(\exposure, \targetexposure) &= \parityreltradeoff\|\exposure\|^2_2 - (1 - \parityreltradeoff)\fdtrans{\exposure}\targetexposure
    \label{eqn:exposure-loss}
\end{align}
where $\targetexposure$ is based on graded relevance (\S\ref{sec:graded}).  
 
Let $\scorer : \docs \rightarrow \Re$ be an item scoring function parameterized by $\scorerparams$.
Given a query, $\score$ is a $\ndocs\times 1$ vector of item scores for the entire collection such that, $\score_{\doc} = \scorer(\doc)$.
Using a Plackett-Luce model, we can translate the raw scores into sampling probabilities,
\begin{align*}
	\prob(\doc) &= \frac{\fdexp{\score_{\doc}}}{\sum_{\doc'\in\docs}\fdexp{\score_{\doc'}}}
\end{align*}
This allows us to construct a ranking $\permutation$ by sampling items sequentially.
Unfortunately, this sampling process is non-differentiable and, therefore, prohibitive to a large class of models, including those that learn by gradient descent.  
We address this by adopting the method proposed by \citet{bruch:stochastic-ltr}.
To construct a sampled ranking $\permutation$, we reparameterize  the probability distribution by adding independently drawn noise samples $\gumbelnoise$ from the Gumbel distribution \citep{maddison2016concrete} to $\score$ and sorting items by the ``noisy'' probability distribution $\probnoise$,
\begin{align}
    \probnoise(\doc_i) &= \frac{\fdexp{\score_{\doc_i} + \gumbelnoise_i}}{\sum_{\doc_j \in \docs}{\fdexp{\score_{\doc_j} + \gumbelnoise_j}}} \label{eqn:noisy-probs}
\end{align}

Given the perturbed probability distribution $\probnoise$, we compute each document's smooth rank \citep{qin2010general, wu2009smoothing} as,
\begin{align}
    \invpermutation_{\doc} &= \sum_{\doc' \in \docs/\doc}\left(1+\fdexp{\frac{\probnoise(\doc) - \probnoise(\doc')}{\temp}}\right)^{-1} \label{eqn:smooth-rank}
\end{align}
The smooth rank is sensitive to the temperature $\temp$. 
At high temperatures the smooth rank is a poor approximation of the true rank and at low temperatures may result in vanishing gradients.
To rectify this issue, we employ the straight-through estimator \citep{bengio2013estimating} to compute the true ranks in forward pass but differentiating the gradients with respect to the smooth ranks during backpropagation.

Using the estimated ranks and a specified user model we compute the exposure for each document.
For example, assuming RBP as the user model the exposure of document $\doc$ from a single ranking $\permutation$ is given by $\exposure_\doc = \rbpp^{\invpermutation_{\doc}}$.
We compute expected exposure by averaging over $\numsamplestrain$ different rankings---each generated by independently sampling different Gumbel noise in Equation \ref{eqn:noisy-probs}.

We use this expected exposure vector $\exposure$ in Equation \ref{eqn:exposure-loss} to compute the loss that we minimize through gradient descent.
The relevance grades are not used for training beyond computing target exposure.
We set $\temp$ in Equation \ref{eqn:smooth-rank} to $0.1$.

We can adapt this model to optimize group-level exposure metrics like demographic parity (\S\ref{sec:fairness}).
To do so, we replace $\|\exposure\|^2_2$ with $\|\iexposure\|^2_2$ in Equation \ref{eqn:exposure-loss} to define an optimization objective that trades-off relevance and demographic parity.
\begin{align}
    \instancelossmodgroup &= \parityreltradeoff\|\iexposure\|^2_2 - (1 - \parityreltradeoff)\fdtrans{\exposure}\targetexposure \label{eqn:exposure-loss-group}
\end{align}
This loss function assumes that the ideal policy distributes exposure equally across all demographics.

\subsection{Experiment}
\label{sec:experiment}

\paragraph{Models}
We restrict our choice of baselines to neural networks so that the exposure-based optimization can be compared to baseline ranking loss functions with respect to the same model.
Our base model consists of a fully-connected neural network with two hidden layers of size 256 nodes per layer and rectified linear unit for activation function.
We choose a learning rate of $0.001$ and a dropout rate of $0.1$ and perform early-stopping for all models based on validation sets.
Stochastic rankings are then derived by employing Plackett-Luce sampling over these deterministic policies (i.e. pointwise and pairwise models), with varying softmax temperatures to obtain different trade-off points between disparity and relevance.
We set $\numsamplestrain$ to 20 for our model and $\numsamplestest$ to 50 for all models.

\paragraph{Objectives}
We consider three models in our experiments.  The pointwise model minimizes the squared error between the model prediction and true relevance.  The pairwise model minimizes misclassified preferences using a cross-entropy loss.  The expected exposure model minimizes the loss in Equation \ref{eqn:exposure-loss} and, in our demographic parity experiments, Equation \ref{eqn:exposure-loss-group}.

\paragraph{Data}
Our experiments use the MSLR-WEB10k dataset \citep{qin2013letor}, a learning-to-rank dataset containing ten thousand queries.
We perform five-fold cross validation ($60/20/20$ split between training, validation, and testing sets).  Each query-document pair is represented by a 136-dimensional feature vector and graded according to relevance on a five point scale.
For the demographic parity experiments, we discretize the PageRank feature in the ranges <1000, 1000--10000, and $\geq$10000 and treat it as a demographic attribute.
We confirm that this discretization scheme is reasonable as roughly $70\%$ of the queries have at least one document corresponding to each demography with a relevance grade greater than one.

\subsection{Results}
\label{sec:optimization-results}

We present the results of our experiments in Table \ref{tbl:results-ltr}.  

\begin{table}[t]
    \centering
    \caption{Results for optimizing towards expected exposure and demographic parity using different ranking objectives.
    We report average $\eeAUC$ for both tasks and highlight the best performance for each in bold.
    Optimizing directly for expected exposure and demographic parity using our proposed method achieves best performance in both cases.}
    \begin{tabular}{lp{1.5cm}p{2.25cm}}
    \hline
    \hline
        \multirow{2}{*}{\textbf{Loss function}} & \multicolumn{2}{c}{\textbf{AUC}} \\
         & \textbf{Expected exposure} & \textbf{Demographic parity} \\
        \hline
        Pointwise loss & $0.229$ & $0.112$ \\
        Pairwise loss & $0.229$ & $0.108$ \\
        \hline
        \multicolumn{2}{l}{\textbf{Our methods}} \\
        Expected exposure  & $\mathbf{0.238}$ & $0.141$ \\
        Demographic parity  & & $\mathbf{0.178}$ \\
        \hline
        \hline
    \end{tabular}
    \label{tbl:results-ltr}
\end{table}

In terms of expected exposure, we did not observe a  difference in performance between pointwise and pairwise models.  However, directly optimizing for expected exposure resulted in a $3.9\%$ improvement  in $\eeAUC$ over the pointwise and pairwise models.
We confirm that the difference in $\eeAUC$ follows a normal distribution and accordingly perform a paired student's t-test to check their statistical significance.
The $\eeAUC$ differences between our proposed method and the baselines are statistically significant $(p < 0.01)$.

In terms of demographic parity, we observe a difference in performance between pointwise and pairwise models.  Moreover, directly optimizing for expected exposure results in improved performance while directly optimizing for demographic parity further boosts performance.
The gap in $\eeAUC$ between all pairs of models are statistically significant $(p < 0.01)$ in this case.

\section{Discussion}
\label{sec:discussion}
Our theoretical results draw clear connections to several areas of information retrieval research.  We believe, moreover, that our empirical results suggest that expected exposure metrics capture important aspects of a retrieval system that are not currently measured in information retrieval evaluation.  Our experiments furthermore demonstrated that these metrics are not only effective for distinguishing systems with varying degrees of expected exposure but also that they can be optimized toward.  

Although previously studied in the context of algorithmic fairness, we have demonstrated that there are deep connections to existing core areas of information retrieval research.  These results warrant revisiting  algorithms and results in classic tasks such as \emph{ad hoc} retrieval, legal search, and diversity-sensitive retrieval.  

Beyond relevance, fairness, and diversity, we believe this approach to evaluation opens avenues for studying probabilistic search systems in probabilistic way.  Many search systems are defined as probabilistic models, capable of handling uncertainty about document relevance \cite{zhu:risky-business}, sometimes using online learning to refine scoring and ranking models and adapt to changing information needs.  These models produce rankings in accordance with a probabilistic policy, so they naturally result in a distribution over rankings associated with each query.  Expected exposure, along with computing expected values of other information retrieval metrics, provides a way to evaluate these models and study the effects of uncertainty.  Moreover, modern search engines also randomize their rankings to reduce bias in feedback data \cite{hofman:thesis}.  Although these systems are often evaluated log data and off-policy evaluation techniques, in the case of pre-launch batch evaluation, we can explicitly model the impact of randomization by evaluating the distribution over rankings.

Randomization and improving equal expected exposure  may also help with user retention.
In search systems, we often want to make sure that we do not overemphasize dominant intents, which can often homogenize populations~\cite{mehrotra:demographics-of-search,liang:fairness-without-demographics}.  As such, randomization can allow us to balance exposure across heterogeneous intents.  Exposure balancing may also prevent churn caused by starvation of producers in two-sided economy systems such as ride-sharing platforms~\cite{suhr:fair-matching-ride-sharing}.

Our exposure model is flexible enough to incorporate more elaborate browsing models.  Several exist others beyond RBP and ERR exist in the literature for rankings which deserve exploration.  Furthermore, as searchers begin to interact with interfaces that are not based on rankings (e.g. two-dimensional grids, three-dimensional environments), alternative user models will need to be developed and incorporated.  

We would also like to note possible limitations of this approach.  First, the impact of randomization on user satisfaction is still an active area of research and we believe cumulative effects of randomization may be a novel extension to explore in the future work \cite{schnabel:exploration}.  Second, from an evaluation perspective, stochastic policies introduce logistical constraints on distribution representation and permutation sampling.  Centralized evaluations like TREC would need to support a method for either interrogating a stochastic policy or requiring a large pool of samples, incurring data storage costs.   Third, although we have focused on randomization in order to increase exposure, we believe that drawing a connection to sequential decision-making scenarios like amortized evaluation are exciting areas of future work.

Notwithstanding these limitations, evaluation through expected exposure, when coupled with stochastic policies, opens a new perspective for the study, understanding, and design of information retrieval systems.

\section{Acknowledgements}
Michael Ekstrand's contribution to this work was supported by the National Science Foundation under Grant No. IIS 17-51278.

\bibliographystyle{ACM-Reference-Format}
\balance
\bibliography{cikm2020}
\end{document}